\documentclass[11pt,showpacs]{revtex4}
\def\comment#1{}

\usepackage
{graphicx}
\begin{document}
\title{The phase and critical point of quantum Einstein-Cartan gravity}
\author{She-Sheng Xue}
\email{xue@icra.it}
\affiliation{
ICRANeT, Piazzale della Repubblica, 10-65122, Pescara, Italy\\
Department of Physics, University of Rome ``La Sapienza'', Piazzale A.~Moro 5, 00185, Rome, Italy
}


\begin{abstract}
By introducing diffeomorphism and local Lorentz gauge invariant holonomy fields, we study in the recent article [S.-S.~Xue, Phys.~Rev.~D82 (2010) 064039] the quantum Einstein-Cartan gravity in the framework of Regge calculus.
On the basis of strong coupling expansion, mean-field approximation and dynamical equations satisfied by holonomy fields, we present in this Letter calculations and discussions to show the phase structure of the quantum Einstein-Cartan gravity, (i) the order phase: long-range condensations of holonomy fields in strong gauge couplings; (ii) the disorder phase: short-range fluctuations of holonomy fields
in weak gauge couplings. According to the competition of the activation energy of holonomy fields and their entropy, we give a simple estimate  of the possible ultra-violet critical point and correlation length for the second-order phase transition from the order phase to disorder one. At this critical point, we discuss whether the continuum field theory of quantum Einstein-Cartan gravity can be possibly approached when the macroscopic correlation length of holonomy field condensations is much larger than the Planck length.

\end{abstract}
\pacs{04.60.Nc,11.10.-z,11.15.Ha,05.30.-d}
\maketitle

\noindent
{\bf Introduction.}
\hskip0.1cm
Since the Regge calculus \cite{regge61,wheeler1964} was proposed for the discretization of gravity theory in 1961, many progresses have been made in the approach of Quantum Regge calculus \cite{hammer_book,hambert2007,shamber}
and its variant dynamical triangulations \cite{loll1999}. In particular,
the renormalization group treatment is applied to 
discuss any possible scale dependence of gravity \cite{hammer_book, hamber_run}. Inspired by the success of 
lattice regularization of non-Abelian gauge theories, the 
gauge-theoretic formulation \cite{smolin1979} of quantum gravity using connection variables on a flat hypercubic lattice of the space-time was studied in the Lagrangian formalism.   
The canonical quantization approaches to the Regge calculus in Hamiltonian formulation are studied in Ref.~\cite{William1986}.  
All these studies are very important steps to understand the Einstein general relativity for gravitational fields in the framework of {\it quantum field theory}.
In our recent articles \cite{xue2009,ec_xue2010}, by introducing diffeomorphism and 
local Lorentz invariant (i.e., {\it local} gauge-invariant) holonomy fields, we present a diffeomorphism and 
local Lorentz invariant regularization and quantization of Euclidean Einstein-Cartan (EC) gravity in the framework of quantum Regge calculus. 

Based on this theoretical formulation of quantum Einstein-Cartan gravity, in this Letter we present a preliminary study of the possible phase structure and ultra-violet critical point for the second-order phase transition of the theory. On the basis of strong coupling expansion, mean-field approximation and dynamical equations satisfied by holonomy fields \cite{xue2009,ec_xue2010}, some calculations and discussions are presented to show the phase structure of the quantum Einstein-Cartan gravity, (i) the order phase: long-range condensations of holonomy fields in strong gauge couplings; (ii) the disorder phase: short-range fluctuations of holonomy fields in weak gauge couplings. Moreover, according to the competition of the activation energy of holonomy fields and their entropy, we give a simple estimate of the possible ultra-violet critical point and correlation length for the second-order phase transition. At this critical point, the minimal area (volume) element is shown to be the Planck one, in addition we discuss whether the sensible continuum field theory of quantum Einstein-Cartan gravity can be possibly approached when the macroscopic correlation length of holonomy field condensations is much larger than the Planck length. The possible relation of this macroscopic correlation length to the cosmological constant scale is also discussed.

\noindent
{\bf Simplicial manifold.}
\hskip0.1cm
The four-dimensional Euclidean manifold ${\mathcal R}^4$
is discretized as an ensemble of ${\mathcal N}_0$ 
space-time points (vertexes) ``${\it x}\in {\mathcal R}^4$'' and ${\mathcal N}_1$ 
links (edges) ``$l_\mu(x)$'' connecting two 
neighboring vertexes.    
The edge (1-simplex) denoted by $(x,\mu)$, connecting two neighboring vertexes labeled by $x$ and $x+a_\mu$  in the forward direction $\mu$, can be represented as a four-vector field $l_\mu(x)$, defined at the vertex ``$x$'' by its forward direction $\mu$ pointing from $x$ to $x+a_\mu$ and its length
\begin{equation}
a_\mu(x)\equiv |l_\mu(x)|\not=0,\quad l_\mu(x)\equiv \tilde a e_\mu(x),
\label{edged0}
\end{equation}
where the fundamental tetrad field $e_\mu(x)\equiv e^a_\mu(x)\gamma_a$ is assigned to each edge (1-simplex) of the simplicial complex, and $\tilde a$ is a characteristic length of the simplicial manifold ${\mathcal M}(\tilde a)$. 
On the edge $(x,\mu)$, we place $U_\mu(x)= e^{ig\tilde a\omega_\mu(x)}$, an 
$SO(4)$ group-valued spin-connection fields $\omega_\mu(x)\equiv \omega^{ab}_\mu(x)\sigma_{ab}$.
The fundamental area operator of the anti-clock like 2-simplex (triangle) $h(x)$ is defined as 
\begin{eqnarray}
S^{\rm h}_{\mu\nu}(x)&\equiv & \,\,
\,\, l_\mu(x)\wedge l^\dagger_{\nu}(x), \quad S_{\rm h}(x)\equiv |S^{\rm h}_{\mu\nu}(x)|.\label{pareaod}
\end{eqnarray}
The fundamental volume element around the vertex ``$x$'' is defined as
\begin{equation}
dV(x)=\sum_{h(x)}dV_h(x),\quad dV_h(x)\equiv S_{\rm h}^2(x)
\label{vold}
\end{equation}
where $\sum_{h(x)}$ indicates the sum over all 2-simplexes $h(x)$ that share the same vertex $x$. 
The characteristic length $\tilde a$ is a running length scale, $\tilde a_1\rightarrow \tilde a_2\cdot\cdot\cdot a_{N-1}\rightarrow\tilde a_N$ and $\tilde a_1> \tilde a_2\cdot\cdot\cdot a_{N-1}>\tilde a_N$, correspondingly simplicial manifold  ${\mathcal M}(\tilde a_1)\rightarrow {\mathcal M}(\tilde a_2)\cdot\cdot\cdot{\mathcal M}(\tilde a_{N-1})\rightarrow{\mathcal M}(\tilde a_N)$. In the sense of Wilson renormalization group invariance, we will try to find a physical scaling region where the macroscopic correlation length $\xi$ of the simplicial manifold is much larger than characteristic length $\tilde a$ that is approaching 
the Planck length $a_{\rm pl}\equiv (8\pi G)^{1/2}$.

\noindent
{\bf Invariant holonomy fields.}
\hskip0.1cm
In Refs.~\cite{xue2009,ec_xue2010}, introducing the vertex field $v_{\mu\nu}(e_\mu,e_\nu)$, we define the diffeomorphism and {\it local} gauge-invariant holonomy field along the loop ${\mathcal C}$ on the Euclidean manifold ${\mathcal R}^4$ 
\begin{eqnarray}
X_{\mathcal C}(v,\omega)&=&
{\mathcal P}_C{\rm tr}\exp\left\{ ig\oint_{\mathcal C}v_{\mu\nu}(x)
\omega^\mu(x) dx^\nu\right\},
\label{pa0s}
\end{eqnarray} 
where ${\mathcal P}_C$ is the path-ordering and ``${\rm tr}$'' denotes the trace over spinor space. 
The regularization of the smallest holonomy field along the closed triangle path of the anti-clock like 2-simplex $h(x)$,
\begin{eqnarray}
X_{h} (v,U)&=& 
{\rm tr}\left[v_{\nu\mu}(x)U_{\mu}(x)v_{\mu\rho}(x+a_\mu)U_{\rho}(x+a_\mu)
v_{\rho\nu}(x+a_\nu)U_{\nu}(x+a_\nu)\right].
\label{xs}
\end{eqnarray}
The diffeomorphism 
and {\it local} gauge-invariant regularized Einstein-Cartan action, 
\begin{eqnarray}
{\mathcal A}_{EC}&=&{\mathcal A}_{P}(g,X_{h})+{\mathcal A}_{H}(\gamma,X_{h}),
\label{ecp}\\
{\mathcal A}_P(e,U)&=&\frac{1}{8g^2}\sum_{h\in {\mathcal M}}
\left[X_{h} (v,U)|_{v_{\mu\nu}(x)  =  e_{\mu\nu}(x)\gamma_5}+{\rm h.c.}\right],
\nonumber\\
{\mathcal A}_H(e,U_\mu)&=&\frac{1}{8g^2\gamma}\sum_{h\in {\mathcal M}}
\left[X_h(v,U)|_{v_{\mu\nu}(x)  =  e_{\mu\nu}(x)}+{\rm h.c.}\right],
\nonumber
\end{eqnarray}  
where $e_{\mu\nu}(x)\equiv \frac{i}{2}\left[e_\mu(x) e_\nu(x)-e_\nu(x) e_\mu(x)\right]$, the Immirzi parameter $\gamma\not=0$ \cite{i1997,rt1998} and
the gauge coupling $g$ depend on the characteristic length $\tilde a$.  
In the naive continuum limit $\tilde a \rightarrow a_{\rm pl}$ and $\tilde ag\omega_\mu\ll 1$:  Eq.~(\ref{ecp}) approaches Einstein-Cartan action [see Section III(F) and Appendix B in Ref.~\cite{ec_xue2010}], when the  running gauge coupling $g(\tilde a)$ satisfies 
\begin{eqnarray}
G_{\rm eff}=\frac{3}{4}\,g(\tilde a)\, \frac{\tilde a^2}{8\pi}\Rightarrow G=\frac{a^2_{\rm pl}}{8\pi}.
\label{effg}
\end{eqnarray}
The partition function and the vacuum expectation value are defined as,
\begin{eqnarray}
Z_{EC}=\int {\mathcal D}e{\mathcal D}U \exp -{\mathcal A}_{EC},\quad
\langle\cdot\cdot\cdot\rangle =\int {\mathcal D}e{\mathcal D}U (\cdot\cdot\cdot)
\exp -{\mathcal A}_{EC}.
\label{par}
\end{eqnarray}
The $\langle X_h\rangle$ obeys the dynamical equation,
\begin{eqnarray}
\langle X_h\rangle =  \langle X_h\Big( 
U_\mu\frac{\delta {\mathcal A}_{EC}}{\delta U_\mu}\Big)\rangle - \langle X_h \Big(U^{\dagger }_\mu\frac{\delta {\mathcal A}_{EC}}{\delta U^{\dagger}_\mu}\Big)\rangle.
\label{sinv2}
\end{eqnarray}
It should be mentioned that if ${\mathcal O}$ is not a diffeomorphism and local Lorentz gauge invariant operator, its vacuum expectation values must vanish $\langle{\mathcal O}\rangle\equiv 0$, because diffeomorphism and local gauge symmetries are exactly preserved without any either explicit or spontaneous breaking.

\noindent
{\bf Mean-field approximation.}
\hskip0.1cm
In order to show the phase structure and transition of regularized Einstein-Cartan theory (\ref{ecp}), one needs to calculate $\langle X_h\rangle$ as a function of the gauge coupling $g$ and the Immirzi parameter $\gamma$. The diffeomorphism and local Lorentz gauge invariant $\langle X_h\rangle$ acts as an order parameter. However, an analytical calculation of $\langle X_h\rangle$ is rather difficult for its non-perturbative nature.
We adopt the mean-field approximation, though it is not
diffeomorphism and local Lorentz gauge invariant, and try to gain some insight into the phase structure and transition of regularized Einstein-Cartan theory.   

In Section VI of Ref.~\cite{ec_xue2010}, introducing the mean-field value $M^2_{\rm h}\sim \langle v^2\rangle$ and averaged area $\langle S_{\rm h}\rangle = M_{\rm h}\tilde a^2$,  for each 2-simplex $h$ 
we define the {\it local} mean-field action $\bar {\mathcal A}_{\rm h}$
for the 2-simplex $h(x)$
\begin{eqnarray}
\bar {\mathcal A}_{\rm h}& = &  {\rm tr}\left[e_\nu(x)\Gamma^h_{\nu\mu}(x)e_\mu(x)-e_\mu(x)\Gamma^h_{\nu\mu}(x)e_\nu(x)\right],
\label{action2d}
\end{eqnarray}
and the {\it local} mean-field partition function
\begin{eqnarray}
\bar Z_{\rm h}=\int_h{\mathcal D}U{\mathcal D}e \exp -\bar{\mathcal A}_{\rm h},\quad \int_h{\mathcal D}U{\mathcal D}e \equiv \int_h dU_\mu dU_\nu dU_\rho de_\mu de_\nu,
\label{meanpar0}
\end{eqnarray}
where 
\begin{eqnarray}
\Gamma^h_{\nu\mu}(x) &=& 
\left(\frac{i}{2}\right)\left(\frac{M^2_{\rm h}}{8g^2}\right)\left(\gamma_5-\frac{1}{\gamma}\right)\left[U_{\nu}(x)U_{\rho}(x+a_\nu)U^\dagger_{\mu}(x)\right]+{\rm h.c.},
\label{chg2t}
\end{eqnarray}
Thus, the regularized EC action
(\ref{ecp}) and partition function (\ref{par}) are approximated by their mean-field counterparts,
\begin{eqnarray}
\bar{\mathcal A}_{EC}=\sum_{h\in {\mathcal M}}\bar{\mathcal A}_{\rm h},\quad \bar Z_{EC}=\prod_{h\in{\mathcal M}}\bar Z_{\rm h}.
\label{meanact}
\end{eqnarray}
Eqs.~(\ref{action2d}-\ref{meanact}) are the mean-field approximation to the regularized Einstein-Cartan theory (\ref{xs},\ref{ecp},\ref{par}). In addition, we adopt the strong coupling expansion in powers of $M^2_{\rm h}/8g^2$ to make analytical calculations. We approximately calculated the free-energy
\begin{eqnarray}
{\mathcal F}_{EC}^{\rm app}(M_{\rm h},g,\gamma, \tilde a) = -{\mathcal N}\ln(1+y_{\rm h}) - {\mathcal N}\frac{2y_{\rm h}}{1+y_{\rm h}}
+\langle {\mathcal A}_{EC}\rangle_\circ, \quad y_{\rm h}\equiv \frac{\gamma^2+1}{64g^4\gamma^2d_j^3} M^4_{\rm h},
\label{appff0}
\end{eqnarray}
where ${\mathcal N}=\sum_{h\in {\mathcal M}}$ is the total number of 2-simplexes and the mean-field value $\langle {\mathcal A}_{EC}\rangle_\circ$ is an average with respect to $\bar Z^h_{EC}$ (\ref{meanact}). By minimizing the free-energy (\ref{appff0}) with respect to $M_{\rm h}$, we try to determine the mean-field value $M^*_{\rm h}(g,\gamma)$, at which the free-energy (\ref{appff0}) reaches its minimum. In Ref.~\cite{ec_xue2010}, the {\it local} mean-field partition function $\bar Z_{\rm h}$ and free-energy (\ref{appff0}) have been obtained in the strong coupling expansion in terms of $M^2_{\rm h}/8g^2$, up to the term 
$(\Gamma^h)^2\sim{\mathcal O}[(M^2_{\rm h}/8g^2)^2]$   
(see Eqs.~(E.5) and (E.6) of Ref.~\cite{ec_xue2010}). 

In the previous article \cite{ec_xue2010}, in order to estimate the minimal area of four-dimensional dynamical simplicial manifold, the term $\langle {\mathcal A}_{EC}\rangle_\circ$ in the free-energy (\ref{appff0}) was very approximately calculated [see Eqs.~(191), (192) and (193) of Ref.~\cite{ec_xue2010}]. In this Letter, in order to gain some sights into the phase structure of regularized Einstein-Cartan theory (\ref{ecp}), we try to improve the calculation $\langle {\mathcal A}_{EC}\rangle_\circ$ in the free-energy (\ref{appff0}) by using strong coupling expansion and the dynamical equation (\ref{sinv2}). This is analogous to the mean-field approach developed \cite{sd_xue1988} for non-perturbative calculations of the Wilson loop in the lattice QCD.

\noindent
{\bf Dynamical equation.}
\hskip0.1cm
We try to solve the dynamical equation for the smallest holonomy fields in the framework of strong coupling expansion and mean-field approximation. 
The vacuum expectation value (\ref{par}) can be written as 
\begin{eqnarray}
\langle \cdot\cdot\cdot\rangle=\frac{\langle (\cdot\cdot\cdot) e^{-({\mathcal A}_{EC}-\bar {\mathcal A}_{EC})}\rangle_\circ}{\langle e^{-({\mathcal A}_{EC}-\bar {\mathcal A}_{EC})}\rangle_\circ} = \langle \cdot\cdot\cdot\rangle_\circ + {\rm hight-order}\,\,{\rm terms},
\label{expaves}
\end{eqnarray} 
where hight-order terms stand for the series of strong coupling expansion ($1/g^2$) of exponential factor $e^{-({\mathcal A}_{EC}-\bar {\mathcal A}_{EC})}$ in both nominator and denominator.
Up to the leading order, replacing $\langle \cdot\cdot\cdot\rangle$ by $\langle \cdot\cdot\cdot\rangle_\circ$ in Eq.~(\ref{sinv2}), we approximately write Eq.~(\ref{sinv2}) as follows, 
\begin{eqnarray}
\langle X_h\rangle &\approx & \langle X_h\rangle_\circ\approx 4g^2\langle{\mathcal A}_{EC}\rangle_\circ/{\mathcal N}\nonumber\\
 \langle {\mathcal A}_{EC}\rangle_\circ &\approx & \langle {\mathcal A}_{EC}\Big( 
U_\mu\frac{\delta {\mathcal A}_{EC}}{\delta U_\mu}\Big)\rangle_\circ - \langle {\mathcal A}_{EC} \Big(U^{\dagger }_\mu\frac{\delta {\mathcal A}_{EC}}{\delta U^{\dagger}_\mu}\Big)\rangle_\circ,
\label{asinv2}
\end{eqnarray}
where the first line bases on Eq.~(\ref{ecp}).
In below, we try to approximately calculate the right-handed side of Eq.~(\ref{asinv2}) to obtain $\langle {\mathcal A}_{EC}\rangle_\circ$ as a function of $M^2_{\rm h}$ and $1/g^2$. Then, substituting the result $\langle {\mathcal A}_{EC}\rangle_\circ=f(M^2_{\rm h},1/g^2)$ into Eq.~(\ref{appff0}), we obtain the approximate free-energy as a function of $M^2_{\rm h}$ and $1/g^2$.

\begin{figure}[ptb]
\includegraphics[scale=1.2
]{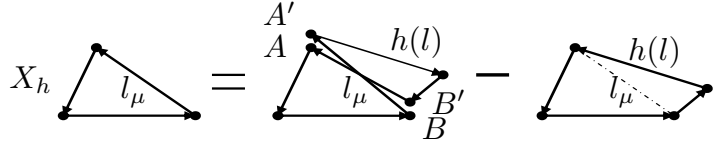}
\comment{
\put(-245,33){$X_h$}
\put(-210,30){$l_\mu$}
\put(-140,30){$l_\mu$}
\put(-56,30){$l_\mu$}
\put(-125,45){$h(l)$}
\put(-50,42){$h(l)$}
\put(-165,43){$A$}
\put(-165,53){$A'$}
\put(-115,17){$B$}
\put(-112,25){$B'$}
}
\caption{We sketch a graphic representation of the dynamical equation (\ref{sinv2}) for the smallest holonomy field $\langle X_h\rangle$ (\ref{xs}). 
Note that $A$ and $A'$ are the same vertex, so are $B$ and $B'$.  
In the right-hand side of the graphic equation, the summation over all 2-simplexes
$h(l)$ associated to this edge $l_\mu$ is made. This figure is reproduced from Fig.~3 in Ref.~\cite{ec_xue2010}.
}%
\label{sdf0}%
\end{figure}
\begin{figure}[tbhp]

\unitlength1mm
\def\fsz{\footnotesize}
\def\ssz{\scriptsize}
\def\tsz{\tiny}
\def\dst{\displaystyle}
\def\pu#1#2{\put(#1,#2){\emmoveto}}
\def\pd#1#2{\put(#1,#2){\emlineto}}
\begin{picture}(105.64,28.645)
\def\dst{\displaystyle}
\def\fsz{\footnotesize}
\put(-3,0){\includegraphics[width=4.6cm]{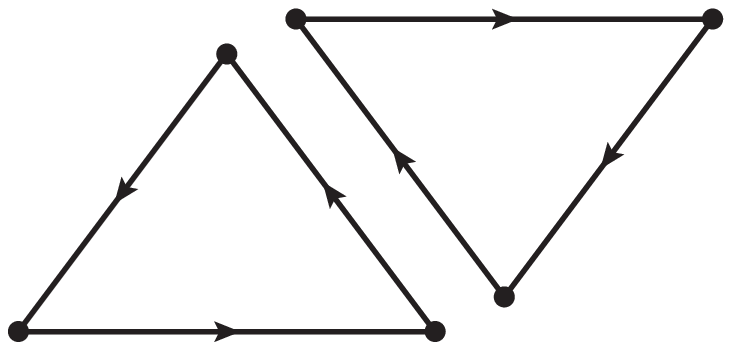}}
\put(12,2){$\mu$}
\put(5,13){$\nu$}
\put(19,12){$\rho$}
\put(12,10){$1$}
\put(25,15){$2$}
\put(25,23){$\nu'$}
\put(32,10){$\mu'$}
\put(40,0){\includegraphics[width=4.6cm]{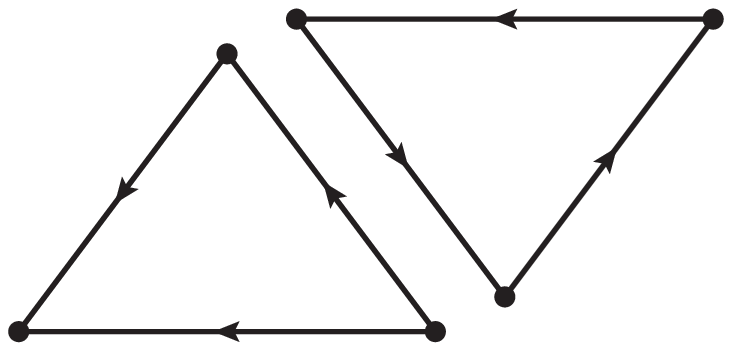}}
\put(55,2){$\mu$}
\put(47,13){$\nu$}
\put(62,12){$\rho$}
\put(55,10){$1$}
\put(69,15){$2$}
\put(68,23){$\nu'$}
\put(75,10){$\mu'$}
\end{picture}
\caption[]{Two simplexes $h_1$ and $h_2$ are not overlap and have a common 1-simplex (link) ``$l$'' in the direction $\rho$. There are $({\mathcal D}_l-1)$ possibilities. Left: this is the second graphic representation in Fig.~\ref{sdf0}. Right:
this is the third graphic representation in Fig.~\ref{sdf0}.}
\label{fig_1_2}
\end{figure}
\begin{figure}[tbhp]

\unitlength1mm
\def\fsz{\footnotesize}
\def\ssz{\scriptsize}
\def\tsz{\tiny}
\def\dst{\displaystyle}
\def\pu#1#2{\put(#1,#2){\emmoveto}}
\def\pd#1#2{\put(#1,#2){\emlineto}}
\begin{picture}(105.64,28.645)
\def\dst{\displaystyle}
\def\fsz{\footnotesize}
\put(-5,0){\includegraphics[width=4.6cm]{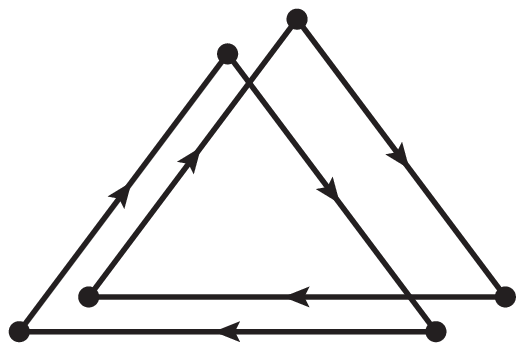}}
\put(15,5){$\mu$}
\put(4,15){$\nu$}
\put(28,18){$\rho$}
\put(15,15){$1$}
\put(25,15){$2$}
\put(45,0){\includegraphics[width=4.6cm]{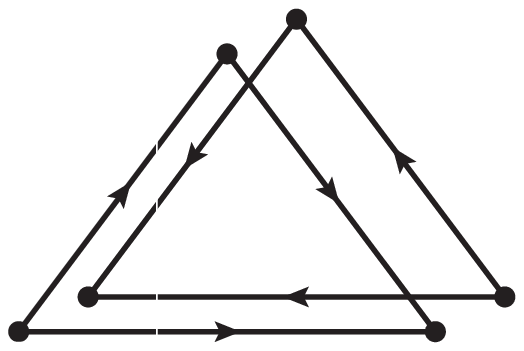}}
\put(65,5){$\mu$}
\put(54,15){$\nu$}
\put(78,18){$\rho$}
\put(65,15){$1$}
\put(73,15){$2$}
\end{picture}
\caption[]{Two simplexes $h_1$ and $h_2$ are completely overlap. Left: this is the second graphic representation in Fig.~\ref{sdf0}.  Right: this is the third graphic representation in Fig.~\ref{sdf0}.}
\label{fig_3_4}
\end{figure}

The graphic representation of Eq.~(\ref{asinv2}) is given in Fig.~\ref{sdf0}. To obtain $\langle {\mathcal A}_{EC}\rangle_\circ$ from Eq.~(\ref{asinv2}), we need to calculate the following four types of diagrams, shown in Figs.~\ref{fig_1_2} and \ref{fig_3_4} .
For the diagrams represented in Fig.~\ref{fig_1_2}, 
using the mean-field approach (\ref{action2d}-\ref{meanact}) and indicating $[\cdot\cdot\cdot]$ to be anti-clock like and $[\cdot\cdot\cdot]^\dagger$ clock like, we have  
\begin{eqnarray}
\langle {\rm Fig}.\ref{fig_1_2}({\rm left})\rangle_\circ &\approx &  \left\{\langle{\rm tr}\Big[e_\nu\Gamma^{h_1}_{\nu\mu}e_\mu-e_\mu\Gamma^{h_1}_{\nu\mu}e_\nu\Big]{\rm tr}\left[e_{\nu'}\Gamma^{h_2}_{\nu'\mu'}e_{\mu'}-e_{\mu'}\Gamma^{h_2}_{\nu'\mu'}e_{\nu'}\right]^\dagger\rangle_\circ\right\}\nonumber\\
&= &\frac{1}{Z_{h_1}Z_{h_2}}\int [dedU]_{h_1}\int [dedU]_{h_2}\,\,\exp -\left[\bar{\mathcal A}_{h_1}+\bar{\mathcal A}_{h_2}\right]
\nonumber\\
&\times& {\rm tr}\Big[e_\nu\Gamma^{h_1}_{\nu\mu}e_\mu-e_\mu\Gamma^{h_1}_{\nu\mu}e_\nu\Big]{\rm tr}\left[e_{\nu'}\Gamma^{h_2}_{\nu'\mu'}e_{\mu'}-e_{\mu'}\Gamma^{h_2}_{\nu'\mu'}e_{\nu'}\right]^\dagger
\nonumber\\
&\approx & 
\langle{\rm tr}\Big[e_\nu\Gamma^{h_1}_{\nu\mu}e_\mu-e_\mu\Gamma^{h_1}_{\nu\mu}e_\nu\Big]\rangle^{h_1}_\circ\langle{\rm tr}\left[e_{\nu'}\Gamma^{h_2}_{\nu'\mu'}e_{\mu'}-e_{\mu'}\Gamma^{h_2}_{\nu'\mu'}e_{\nu'}\right]^\dagger\rangle^{h_2}_\circ\nonumber\\
& = &\sum_{h\in {\mathcal M}} ({\mathcal D}_l-1)C_r
\left(\frac{y_{\rm h}}{\bar Z_{\rm h}}\right)^2,
\label{afao31}
\end{eqnarray}
where in the last line we use
Eqs.~(E1)-(E7) in Appendix E of Ref.~\cite{ec_xue2010}. Due to the relations between anti-clock like orientation and clock like orientation $\Gamma^{h}_{\nu\mu}=\Gamma^{h\dagger }_{\nu\mu}$ and $e_\nu\Gamma^{h}_{\nu\mu}e_\mu=-e_\mu\Gamma^{h}_{\nu\mu}e_\nu$ (see Fig.~\ref{fig_1_2}), we have
\begin{eqnarray}
\langle {\rm Fig}.\ref{fig_1_2}({\rm right})\rangle_\circ
&\approx &  \left\{\langle{\rm tr}\Big[e_\nu\Gamma^{h_1}_{\nu\mu}e_\mu-e_\mu\Gamma^{h_1}_{\nu\mu}e_\nu\Big]{\rm tr}\left[e_{\nu'}\Gamma^{h_2}_{\nu'\mu'}e_{\mu'}-e_{\mu'}\Gamma^{h_2}_{\nu'\mu'}e_{\nu'}\right]\rangle_\circ\right\}
\nonumber\\
&\approx & -
\langle{\rm tr}\Big[e_\nu\Gamma^{h_1}_{\nu\mu}e_\mu-e_\mu\Gamma^{h_1}_{\nu\mu}e_\nu\Big]\rangle^{h_1}_\circ\langle{\rm tr}\left[e_{\nu'}\Gamma^{h_2}_{\nu'\mu'}e_{\mu'}-e_{\mu'}\Gamma^{h_2}_{\nu'\mu'}e_{\nu'}\right]^\dagger\rangle^{h_2}_\circ\nonumber\\
& = &-\sum_{h\in {\mathcal M}} ({\mathcal D}_l-1)C_r
\left(\frac{y_{\rm h}}{\bar Z_{\rm h}}\right)^2,
\label{afao31_2}
\end{eqnarray}
where in the last line we use Eq.~(\ref{afao31}). For these two cases,
$\mu\not=\mu'$, $\nu\not=\nu'$ and
\begin{eqnarray}
C_r\equiv \Big[\Big(\frac{\gamma^2+1}{\gamma^2}\Big)^2+\frac{4}{\gamma^2}\Big]
\Big(\frac{\gamma^2+1}{\gamma^2}\Big)^{-2},
\label{c_r}
\end{eqnarray}
and ${\mathcal D}_l$
is the total number of 2-simplexes $h$ associating the fixed link $l$ \footnote{In the case of a $d$-dimensional hypercubic lattice, the total number of plaquettes associating the fixed link $l$ is ${\mathcal D}_l=2(d-1)$}. 
For the diagrams represented in Fig.~\ref{fig_3_4},
we have  
\begin{eqnarray}
\langle {\rm Fig}.\ref{fig_3_4}({\rm left})\rangle_\circ &\approx &  \left\{\langle{\rm tr}\Big[e_\nu\Gamma^{h_1}_{\nu\mu}e_\mu-e_\mu\Gamma^{h_1}_{\nu\mu}e_\nu\Big]{\rm tr}\left[e_{\nu}\Gamma^{h_1}_{\nu\mu}e_{\mu}-e_{\mu}\Gamma^{h_1}_{\nu\mu}e_{\nu}\right]\rangle_\circ\right\}\nonumber\\
&= &\frac{1}{Z_{h_1}}\int [dedU]_{h_1}\,\,\exp -\left[\bar{\mathcal A}_{h_1}\right]
\nonumber\\
&\times& {\rm tr}\Big[e_\nu\Gamma^{h_1}_{\nu\mu}e_\mu-e_\mu\Gamma^{h_1}_{\nu\mu}e_\nu\Big]{\rm tr}\left[e_{\nu}\Gamma^{h_1}_{\nu\mu}e_{\mu}-e_{\mu}\Gamma^{h_1}_{\nu\mu}e_{\nu}\right]
\nonumber\\
&= & 
\frac{4}{Z_{h_1}}\int [dedU]_{h_1} \Big[{\rm tr}(e_\nu\Gamma^{h_1}_{\nu\mu}e_\mu)\Big]^2\,\exp -\left[\bar{\mathcal A}_{h_1}\right]
\nonumber\\
& \approx  & \frac{4}{Z_{h_1}}\int [dU_\mu dU_\nu dU_\rho]\left(\frac{3}{4}\right)
\frac{\Gamma^{h_1}_{\nu\mu}\Gamma^{h_1}_{\nu\mu}}{(1-\Gamma^{h_1}_{\nu\mu})^2} 
\det^{\hskip0.5cm -1}[1-\Gamma^{h_1}_{\nu\mu}]
\label{ainte4}\\
& \approx & 3\sum_{h\in {\mathcal M}} C_r
\left(\frac{y_{\rm h}^2}{\bar Z_{\rm h}}\right),
\label{afao32}
\end{eqnarray}
and
\begin{eqnarray}
\langle {\rm Fig}.\ref{fig_3_4}({\rm right})\rangle_\circ &\approx &  \left\{\langle{\rm tr}\Big[e_\nu\Gamma^{h_1}_{\nu\mu}e_\mu-e_\mu\Gamma^{h_1}_{\nu\mu}e_\nu\Big]{\rm tr}\left[e_{\nu}\Gamma^{h_1}_{\nu\mu}e_{\mu}-e_{\mu}\Gamma^{h_1}_{\nu\mu}e_{\nu}\right]^\dagger\rangle_\circ\right\}\nonumber\\
&= &\frac{1}{Z_{h_1}}\int [dedU]_{h_1}\,\,\exp -\left[\bar{\mathcal A}_{h_1}\right]
\nonumber\\
&\times& {\rm tr}\Big[e_\nu\Gamma^{h_1}_{\nu\mu}e_\mu-e_\mu\Gamma^{h_1}_{\nu\mu}e_\nu\Big]{\rm tr}\left[e_{\nu}\Gamma^{h_1}_{\nu\mu}e_{\mu}-e_{\mu}\Gamma^{h_1}_{\nu\mu}e_{\nu}\right]^\dagger
\nonumber\\
&= & 
\frac{4}{Z_{h_1}}\int [dedU]_{h_1} \Big[{\rm tr}(e_\nu\Gamma^{h_1}_{\nu\mu}e_\mu)\Big]\Big[{\rm tr}(e_\nu\Gamma^{h_1}_{\nu\mu}e_\mu)\Big]^\dagger\,\exp -\left[\bar{\mathcal A}_{h_1}\right]
\nonumber\\
& \approx  & -\frac{4}{Z_{h_1}}\int [dU_\mu dU_\nu dU_\rho]
\frac{\Gamma^{h_1}_{\nu\mu}[\Gamma^{h_1}_{\nu\mu}]^\dagger}{(1-\Gamma^{h_1}_{\nu\mu})^2} 
\det^{\hskip0.5cm -1}[1-\Gamma^{h_1}_{\nu\mu}]
\label{ainte5}\\
& \approx & -2\sum_{h\in {\mathcal M}}
\left(\frac{y_{\rm h}}{\bar Z_{\rm h}}\right).
\label{afao33}
\end{eqnarray}
In lines (\ref{afao32}) and (\ref{afao33}), we use Eqs.~(183)-(185) and
Eqs.~(E5)-(E6) in Appendix E of Ref.~\cite{ec_xue2010}, and  in the lines (\ref{ainte4}) and (\ref{ainte5}), we use 
\begin{eqnarray}
\int_h d e_\mu de_\nu \,e_{\mu}^2e_{\nu}^2\, \exp -\bar{\mathcal A_{\rm h}}
&=&\frac{3}{4}\Big\{[I-\Gamma^h]^{-2}_{\mu\nu}\Big\}{\rm det}^{-1}[I-\Gamma^h].
\label{det4}
\end{eqnarray}
As a result, we obtain the $\langle {\mathcal A}_{EC}\rangle_\circ$ as a function of $M^2_{\rm h}$  and $1/g^2$,
\begin{eqnarray}
\langle {\mathcal A}_{EC}\rangle_\circ\approx {\mathcal N}\left\{ 2({\mathcal D}_l-1)C_r
\left(\frac{y_{\rm h}}{\bar Z_{\rm h}}\right)^2 + 3C_r\left(\frac{y_{\rm h}^2}{\bar Z_{\rm h}}\right)+2
\left(\frac{y_{\rm h}}{\bar Z_{\rm h}}\right)\right\}.
\label{dxsol}
\end{eqnarray}

%

\begin{figure}[ptb]
\includegraphics[scale=0.65]{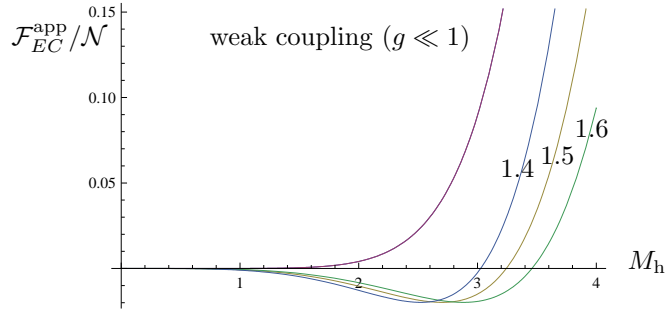}
\put(8,15){$M_{\rm h}$}
\put(-150,100){weak coupling $(g\ll 1)$}
\put(-12,65){$1.6$}
\put(-25,55){$1.5$}
\put(-40,50){$1.4$}
\put(-225,100){${\mathcal F}^{\rm app}_{EC}/{\mathcal N}$}
\caption{In the strong coupling region $(g> 1)$, the approximate free-energy (\ref{appff1}) is plotted as a function of $M_{\rm h}$ $(\tilde a=1)$.   
It shows that the location of minimal free-energy $M^*_{\rm h}(g)$ becomes small, as the gauge coupling decreases, i.e.,
$M^*_{\rm h}(g_1)< M^*_{\rm h}(g_2)< M^*_{\rm h}(g_3)$ for $g_1=1.4$, $g_2=1.5$ and $g_3=1.6$. In the weak gauge coupling region $(g\ll 1)$,
a speculated free-energy with its minimum at $M^*_{\rm h}=0$ is sketched.  
We approximately adopt ${\mathcal D}_l\approx 6$ in Eq.~(\ref{dxsol}) for a four-dimensional simplicial manifold, and results are not sensitive to ${\mathcal D}_l$ values.  }%
\label{free-m}%
\end{figure}

\noindent
{\bf free-energy and two-phase structure.}
\hskip0.1cm
Substituting Eq.~(\ref{dxsol}) into Eqs.~(\ref{appff0}) , we obtain the approximate free-energy 
\begin{eqnarray}
\frac{1}{{\mathcal N}}{\mathcal F}_{EC}^{\rm app}(M_{\rm h},g,\gamma, \tilde a) = -\ln(1+y_{\rm h}) 
+2({\mathcal D}_l-1)C_r
\left(\frac{y_{\rm h}}{\bar Z_{\rm h}}\right)^2 + 3C_r\left(\frac{y_{\rm h}^2}{\bar Z_{\rm h}}\right).
\label{appff1}
\end{eqnarray}
Its minimum, as shown in Fig.~\ref{free-m}, locates at
\begin{eqnarray}
M^*_{\rm h}\approx y^{\rm min}_{\rm h}\left(\frac{2\gamma^2d_j^3}{\gamma^2+1}\right)^{1/2}(2g)\approx 0.91 g, 
\label{dxsol1}
\end{eqnarray}
where $y^{\rm min}_{\rm h}\approx 0.04$, the Immirzi parameter $\gamma \gg 1$ and fundamental representation $d_j=4$. The result (\ref{dxsol1}) shows a nonvanishing mean-field value $M^*_{\rm h}$ decreases as the gauge coupling $g$ decreases. The mean-field value $M^*_{\rm h}$ (\ref{dxsol1}) shows that the framework of mean-field approximation and strong couping expansion is self-consistent for the expanding parameter $(M^*_{\rm h})^2/(8g^2)\approx 0.1$ being smaller than one.

Submitting $\,y^{\rm min}_{\rm h}\approx 0.04\,$, the location of free-energy minimum,  into Eq.~(\ref{dxsol}), we obtain the vacuum expectation value
\begin{eqnarray}
\langle {\mathcal A}_{EC}\rangle/{\mathcal N}\approx\langle {\mathcal A}_{EC}\rangle_\circ/{\mathcal N}\approx 0.1.
\label{eveaec}
\end{eqnarray}
From the first line of Eqs.~(\ref{ecp}), we approximately have 
\begin{eqnarray}
\langle X_h\rangle\approx\langle X_h\rangle_\circ\approx 4g^2\langle{\mathcal A}_{EC}\rangle_\circ/{\mathcal N}\simeq 0.4 g^2.\nonumber
\end{eqnarray} 
The mean-field values for the 2-simplex area (\ref{pareaod}) and 
the volume element (\ref{vold}) are 
\begin{eqnarray}
\langle S_{\rm h}(x) \rangle = \tilde a^2 M^*_{\rm h}\approx 0.91\,\, g\,\tilde a^2,\quad
\langle dV(x) \rangle = \tilde a^4 N_{\rm h}(M^*_{\rm h})^2
\approx 0.83 \,g^2 \tilde a^4 N_{\rm h} ,
\label{aovm}
\end{eqnarray} 
where $N_{\rm h}$ is the mean value of the number of 2-simplexes $h(x)$ that share the same vertex. 
These nonvanishing values (\ref{dxsol1}-\ref{aovm}) characterize an order phase in strong gauge couplings, as will be discussed below.

Henceforth, Eqs.~(\ref{pa0s},\ref{xs}) will be called $X$-loop for short. The regularized Einstein-Cartan action ${\mathcal A}_{EC}$ is actually the ratio of the activation energy per area (the smallest  
$X$-loop $X_h$) and squared gauge coupling $g^2$ (``temperature''). In the order phase ($\langle X_h\rangle\not= 0$ or $M^*_{\rm h}\not= 0$) for large coupling $g\gg 1$, 
the $X$-loops (\ref{pa0s}) are not suppressed because the smallest $X$-loops $X_h$ undergo condensation by jointing together side by side to form surfaces whose boundaries appears as large $X$-loops. Namely, these $X$-loops proliferate and become macroscopic in the length $\xi$ that is the coherence correlation length of the system, leading to the area law   
\begin{eqnarray}
\langle X_{\mathcal C}\rangle &\sim & (\langle X_h\rangle)^n \sim \exp [- A_{\rm min}({\mathcal C})/\langle S_{\rm h}(x) \rangle],
\label{xarea0}
\end{eqnarray}
where $n=A_{\rm min}({\mathcal C})/\langle S_{\rm h}(x) \rangle$ is the minimal number of 2-simplexes filling the minimal area $A_{\rm min}({\mathcal C})$ that can be spanned by the loop ${\mathcal C}$ and $\langle S_{\rm h}(x) \rangle \approx \tilde a^2 M^*_{\rm h}$ is the averaged area of 2-simplexes. 

The result (\ref{dxsol1}) is obtained by solving the dynamical equation (\ref{asinv2}) in the framework of mean-field approximation and strong coupling expansion for $(M^*_{\rm h})^2/(8g^2)\ll 1$.  Therefore, this result (\ref{dxsol1}) does not apply to the weak gauge coupling region $g\ll 1$, so that we cannot conclude $M^*_{\rm h}\rightarrow 0$ as $g\rightarrow 0$.
We have not so far been able to do any analytical calculation in the weak gauge coupling region of the regularized Einstein-Cartan action (\ref{ecp}). Nevertheless, we can gain some insight into the possible phase in the weak coupling region by looking at the limit of gauge weak coupling $g\rightarrow 0$ of the regularized Einstein-Cartan action (\ref{ecp}).  
In the limit of weak gauge coupling $g\rightarrow 0$, as we can see from ${\mathcal A}_P$ and ${\mathcal A}_H$, the configurations of tetrad fields $\{e_\mu(x)\}$ and gauge fields $\{U_\mu(x)\}$ have to be frozen to the configurations of small fluctuating fields for 
small $X_h$,
otherwise the partition function (\ref{par}) would vanish. Namely, tetrad fields $\{e_\mu(x)\}$ and gauge fields $\{U_\mu(x)\}$ undergo fluctuations at small scale $\tilde a$ with large entropy. The smallest $X$-loops $X_h$ are suppressed by their activation energy, and become then irrelevant for the large scale behavior of the system. Therefore, in the weak coupling region $g\ll 1$, we conjecture the existence of the disorder phase with $\langle X_h\rangle=0$. In the framework of mean-field approximation, this means that 
the minimum of the free-energy locates at $M^*_{\rm h}=0$,
as sketched in Fig.~\ref{free-m}.

\begin{figure}[ptb]
\includegraphics[scale=0.65]{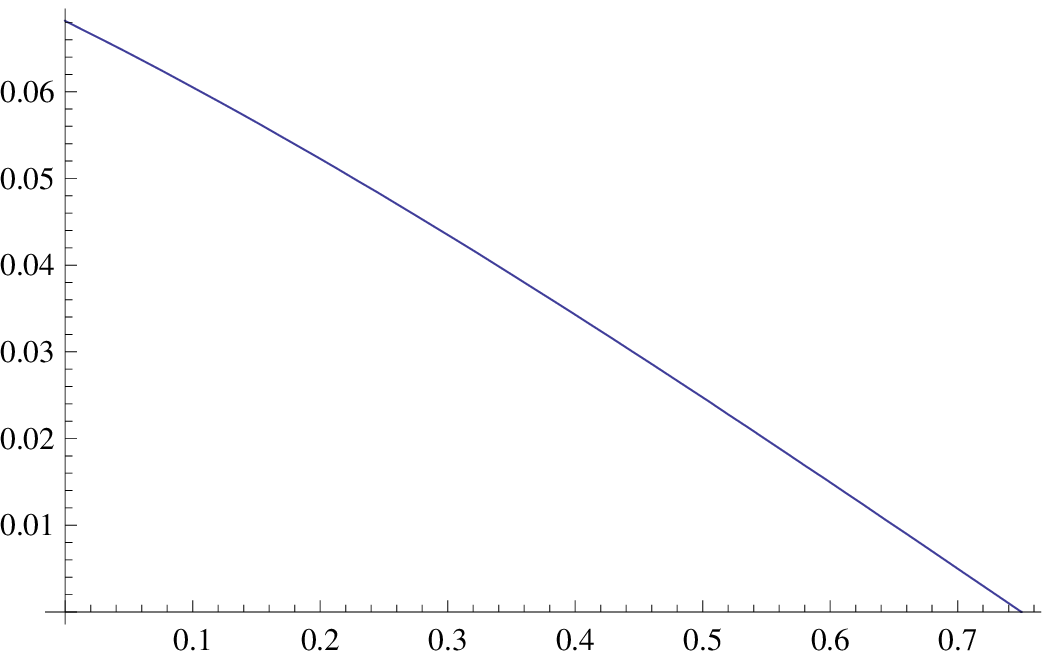}
\put(10,10){$1/g$}
\put(-12,0){$1/g_c$}
\put(-200,120){$1/\gamma$}
\put(-150,60){$M^*_{\rm h}\not=0$}
\put(-150,40){$\langle X_h\rangle\not=0$}
\put(-100,100){$M^*_{\rm h}=0$}
\put(-100,80){$\langle X_h\rangle=0$}
\caption{We sketch the conjectured phase diagram in terms of inverse gauge coupling $1/g$ and Immirzi parameter $1/\gamma$. As discussed in the text, the critical point indicated $g_c=4/3$ for $1/\gamma =0$. While the critical line and point $1/g=0$ and $1/\gamma \simeq 0.068$ are arbitrarily sketched for indicating two-phases structure.   
}%
\label{phase}%
\end{figure}

These two distinct phases in the strong and weak coupling regions are characterized by the order parameter $\langle X_h\rangle$ or 
the mean-field value $M^*_{\rm h}$ in the framework of mean-field approximation. 
Taking into account the Immirzi parameter $\gamma\ge 1$ in the action ${\mathcal A}_H$ [see Eq.~(\ref{ecp})], we find from Eq.~(\ref{chg2t}) that the increasing value of Immirzi parameter $\gamma$ effectively leads the decreasing of gauge coupling $g$. Therefore we conjecture that the phase diagram should be the one sketched in Fig.~\ref{phase}. 
\comment{
As a result the {\it local} mean-field action $\bar {\mathcal A}_{\rm h}$
(\ref{action2d},\ref{chg2t}) becomes
\begin{eqnarray}
\bar {\mathcal A}_{\rm h}&=&{\rm tr}\left[e_\nu(x)\Gamma^h_{\nu\mu}(x)e_\mu(x)-e_\mu(x)\Gamma^h_{\nu\mu}(x)e_\nu(x)\right],
\label{action2dm}\\
\Gamma^h_{\nu\mu}(x) 
&=& \frac{M^{*2}_{\rm h}}{8g^2}
\left(\frac{i}{2}\right)\gamma_5\left[U_{\nu}(x)U_{\rho}(x+a_\nu)U^\dagger_{\mu}(x)\right]+{\rm h.c.},
\nonumber
\end{eqnarray}
where $M^{*2}_{\rm h}/8g^2\approx 0.1$ can be considered as a small coupling of the {\it local} mean-field action $\bar {\mathcal A}_{\rm h}$, and this corresponds to a strong coupling $(g\gg 1)$ of the regularized Einstein-Cartan action (\ref{pact},\ref{diract}). In the order phase, we can use the mean-field action (\ref{action2d}) with the value $M^{*2}_{\rm h}/8g^2\approx 0.1$ to calculate mean-field vacuum expectation values 
\begin{eqnarray}
\langle \cdot\cdot\cdot \rangle^h_\circ &= & \frac{1}{\bar Z_{\rm h}}\int_h{\mathcal D}U{\mathcal D}e \,(\cdot\cdot\cdot)\, e^{-\bar{\mathcal A}_{\rm h}}.
\label{aves}
\end{eqnarray}
to approximate true vacuum expectation values $\langle\cdot\cdot\cdot\rangle$.
}  

\noindent
{\bf Phase transition and critical coupling.}
\hskip0.1cm
Since we have not so far been able to calculate the order parameter $\langle X_h\rangle$ in the weak coupling region, we cannot exactly determine the critical point or line of the second-order phase transition from the order phase to disorder phase. Nevertheless, in this section, we try to discuss  the critical point or line of the second-order phase transition. 

As indicated in Fig.~\ref{free-m}, for the disorder phase in the weak coupling region, the minimal free-energy is zero locating at $\langle X_h\rangle=0$; for the order phase in the strong coupling region, the minimal free-energy is negative locating at $\langle X_h\rangle\not=0$. The second-order phase transition from the order phase to disorder phase occurs when the minimum of free-energy for weak couplings flips into the one for strong couplings, as a result of the competition between the activation energy of $X$-loops and their entropy.  
We expect the second-order phase transition taking place at a critical coupling $g=g_c\not=0$ for $\gamma \gg 1$, and we try to estimate it in a simple way that was adopted to estimate the critical point of the second-order phase transition for two-dimensional systems \cite{kt}, the superfluid helium \cite{kleinert} and the $U(1)$ lattice gauge theory \cite{kogut}. 

For this purpose, we consider the partition function of a single $X$-loop of arbitrary length ${\mathcal C}$ given by the integral
\begin{eqnarray}
Z_{\mathcal C}&=&\int {\mathcal D}e{\mathcal D}U (X_{\mathcal C})\exp -{\mathcal A}_{EC}\nonumber\\
&\sim& \int {\mathcal D}A({\mathcal C}) (2d)^{A({\mathcal C})/\langle S_{\rm h}\rangle}\exp -\frac{1}{(2g)^2}\sum_{h\in A({\mathcal C})}\langle X_h\rangle,
\label{loop-par}
\end{eqnarray}
where (i) $\int {\mathcal D}A({\mathcal C})$ is the functional measure of all possible surface-area $A({\mathcal C})$ bound by the closed loop ${\mathcal C}$; (ii) $(2d)^{A({\mathcal C})/\langle S_{\rm h}\rangle}$ approximately accounts for the number of possible configurations (surface deformations at the area scale $\langle S_{\rm h}\rangle$) of the surface with a given surface-area $A({\mathcal C})$, this is related to the entropy of the surface-area $A({\mathcal C})$; (iii) $\langle X_h\rangle$ is the activation energy of a 2-simplex surface-area $\langle S_{\rm h}\rangle$, $\sum_{h\in A({\mathcal C})}\langle X_h\rangle$ stands for the activation energy of the surface-area $A({\mathcal C})$ and $(2g)^2$ plays a role of ``temperature''. The number of local deformations of a 2-simplex area in a three-dimensional simplicial manifold  is 6, and the number of local deformations of a 1-simplex length in a two-dimensional simplicial manifold 4. We assume this number to be $2d$ in a $d$-dimensional simplicial manifold.  This assumption is not crucial, as you will see below, for a simple estimate of the critical coupling. The free-energy of a grand-canonical ensemble of arbitrary surface-area $A({\mathcal C})$ of the loop ${\mathcal C}$ is then given by $F_{\mathcal C}\sim -\ln Z_{\mathcal C}$, and  
\begin{eqnarray}
Z_{\mathcal C}
&\sim& \int {\mathcal D}A({\mathcal C}) \,\,\,\exp \,\,\,+\,\,\frac{A({\mathcal C})}{\langle S_{\rm h}\rangle}\ln (2d) -\frac{1}{(2g)^2}\sum_{h\in A({\mathcal C})}\langle X_h\rangle\nonumber\\
&\simeq& \int {\mathcal D}A({\mathcal C}) \,\,\,\exp \,\,\,+\frac{A({\mathcal C})}{\langle S_{\rm h}\rangle}\left[\ln (2d) -\frac{1}{(2g)^2}\langle X_h\rangle\right].
\label{loop-par1}
\end{eqnarray}
This integral converges only below a critical gauge coupling $g_c$
\begin{eqnarray}
g_c =\frac{1}{2} \left[\frac{\langle X_h\rangle}{\ln (2d)}\right]^{1/2}.
\label{critical}
\end{eqnarray}
At the critical coupling $g_c$, a single 2-simplex configuration ($X_h$) is activated and its activation energy should be the order of the Planck scale $1/a_{\rm pl}$, i.e., $\langle X_h\rangle \sim {\mathcal O}(1)$, as the characteristic length of simplicial manifold is approaching the Planck length $(\tilde a\rightarrow a_{\rm pl})$. Eq.~(\ref{critical}) gives the critical coupling  $g_c\sim {\mathcal O}(1)$.
  
Above the critical coupling $g_c$, the integral diverges and the ensemble undergoes the second-order phase transition in which the surface-area proliferates and
becomes macroscopically large with the coherent correlation length $\xi\gg \tilde a$. This means that the surface-area $A({\mathcal C})$ of an $X$-loop ($X_{\mathcal C}$) can only be easily deformed beyond the scale $\xi^2$, indicating the ``{\it condensation of $X$-loops }'', and the ensemble stays in the order phase. 
\comment{
and the number of possible configurations follows the area-law
\begin{eqnarray}
(2d)^{A({\mathcal C})/\xi^2}=\exp [ \frac{A({\mathcal C})}{\xi^2}\ln (2d)],
\label{nco}
\end{eqnarray}
}
The divergent integral (\ref{loop-par1}) can be written as 
\begin{eqnarray}
Z_{\mathcal C}
&\sim&  \int {\mathcal D}A({\mathcal C}) \,\,\,\exp \,\,\,+\,\,A({\mathcal C})/\xi^2,\nonumber\\
\xi^2 &\equiv & \frac{\langle S_{\rm h}\rangle}{\ln (2d)}\cdot\left(\frac{g^2}{g^2-g_c^2}\right)
\label{loop-par2}
\end{eqnarray}
for $g>g_c$.
Below the critical coupling $g_c$, none of 2-simplex configurations ($X_h$) is activated and $\langle X_h \rangle=0$, there are short distance ($\sim \tilde a$) fluctuations of tetrad fields $e_\mu(x)$ and group-valued fields $U_\mu(x)$ with large entropy, and the ensemble stays in the disorder phase. This implies that the second-order phase transition from the order phase to disorder phase takes place at the critical gauge coupling $g_c$. 

In the order phase, as the characteristic length of simplicial manifold is getting smaller and approaching the Planck length $(\tilde a\rightarrow a_{\rm pl})$, the gauge 
coupling $g(\tilde a)$ is approaching to the critical gauge coupling $g_c$, which is an ultra-violet fix point. The coherent correlation length $\xi$ (\ref{loop-par2}) becomes macroscopically large
\begin{eqnarray}
\xi &=& \left[\frac{\langle S_{\rm h}\rangle}{2\ln (2d)}\right]^{1/2}\cdot\frac{g_c^{1/2}}{(g-g_c)^{1/2}}\nonumber\\
&\approx & 0.48 \cdot\frac{g_c\tilde a}{(g-g_c)^{1/2}}\gg \tilde a,
\label{loop-par3}
\end{eqnarray}
in the neighborhood of the critical coupling $g\sim g_c+0^+$, where 
a {\it quantum field theory} of the Euclidean Einstein-Cartan gravity can  possibly be realized. In the second line of Eq.~(\ref{loop-par3}),
$d=4$ and the mean-field value of 2-simplex area (\ref{aovm}) are used. 
In Eq.~(\ref{loop-par3}), the critical coupling $g_c\sim {\mathcal O}(1)$, critical exponent $\nu=1/2$ and proportional coefficient $c_0\sim 0.48$ are preliminary results obtained in this simple estimation.

In the mean-field approximation we have $\langle X_h\rangle\sim (M^*_{\rm h})^2$.  $\langle X_h\rangle \sim {\mathcal O}(1)$ implies $M^*_{\rm h}\sim {\mathcal O}(1)$ in the nontrivial continuum limit (\ref{loop-par3})
for $\tilde a\rightarrow a_{\rm pl}$ and $g\rightarrow g_c$. 
Therefore,    
the minimal averaged area of 2-simplexes in the mean-field approximation is given by Eqs.~(\ref{dxsol1}) and (\ref{aovm})
\begin{eqnarray}
\langle S_{\rm h}(\tilde a) \rangle &=& \tilde a^2 M^*_{\rm h}=0.91 g(\tilde a)\tilde a^2\Big|_{\tilde a\rightarrow a_{\rm pl},\, g\rightarrow g_c} \sim {\mathcal O}(a^2_{\rm pl})
\label{aov1}
\end{eqnarray} 
which shows the minimal area (volume) element of the space-time is the order of the Planck scale in the nontrivial continuum limit (\ref{loop-par3}), the basic arena of physical reality we live on \cite{preparata91}.
  
In addition to the case in the nontrivial continuum limit (\ref{loop-par3}), we try to discuss this critical coupling $g_c$ by looking at the naive continuum limit of the regularized Einstein-Cartan action (\ref{ecp}), where
the gauge coupling $g=g(\tilde a)$ depends on the characteristic spacing of simplicial manifold $\tilde a$.  In the naive continuum limit $\tilde a\rightarrow a_{\rm pl}$ and $g\tilde a \omega_\mu \ll 1$, we obtain  the effective Newton constant [see Eq.~(\ref{effg})], 
\begin{eqnarray}
G_{\rm eff}(\tilde a)=\frac{3}{4}\,g(\tilde a)\, \frac{\tilde a^2}{8\pi},
\label{effg1}
\end{eqnarray}
which has to approach to the Newton constant $G=a_{\rm pl}^2/(8\pi)$ in the continuum Einstein-Cartan theory. This leads to $g(\tilde a)|_{\tilde a\rightarrow a_{\rm pl}}\rightarrow 4/3+0^+$ and the critical coupling $g_c=4/3$. 

We turn now to a general discussion. As the running gauge coupling $g(\tilde a)$ is approaching to its ultra-violet critical point $g_c$ ($g\rightarrow g_c$) for $\tilde a\rightarrow a_{\rm pl}$, 
physical and dimensionful quantities $m(g,\tilde a)$ should satisfy the renormalization group invariant equation, 
\begin{eqnarray}
\tilde a\frac{d m}{d \tilde a}=\tilde a\frac{\partial m}{\partial \tilde a}-\beta(g)\frac{\partial m}{\partial g}=0,\quad \beta(g)\equiv -\tilde a\frac{\partial g(\tilde a)}{\partial \tilde a},
\label{regeq}
\end{eqnarray}
because the coherent correlation length $\xi$, the physical scale, becomes much larger than $\tilde a$ ($\xi\gg \tilde a$). The running gauge coupling $g(\tilde a/\xi) > g_c$  can be expanded as a series ,
\begin{eqnarray}
g(\tilde a/\xi)&=& g_c \left[1+ a_0(\tilde a/\xi)^{1/\nu}+ {\mathcal O}[(\tilde a/\xi)^{2/\nu}]\right],
\label{gexp}
\end{eqnarray}
leading to the $\beta$-function 
\begin{eqnarray}
\beta(g)\equiv -\tilde a\frac{\partial g(\tilde a/\xi)}{\partial \tilde a}&=& \beta_0 + \beta_1 (g-g_c) + {\mathcal O}[(g-g_c)^2],
\label{betaexp}
\end{eqnarray}
where $\beta_0=0$ and $\beta_1=-1/\nu$. Assuming $m=\xi^{-1}$, as the solution to the renormalization group invariant equation (\ref{regeq}), one obtains that the coherent correlation length $\xi$ follows the scaling law
\begin{eqnarray}
\xi &= & c_0\tilde a\exp \int^g\frac{dg'}{\beta(g')}=\frac{ c_0\tilde a}{(g-g_c)^\nu},
\label{xivary}
\end{eqnarray}
where the proportional coefficient $c_0=(a_0g_c)^\nu$ and critical exponent $\nu$. Eq.~(\ref{xivary}) has the same form as Eq.~(\ref{loop-par3}). 
Non-perturbative calculations by 
numerical simulations are required to determine the proportional coefficient $c_0$ and the critical exponent $\nu$ in Eq.~(\ref{loop-par3}) or (\ref{xivary}). 
  
\noindent
{\bf Some remarks.}
\hskip0.1cm
In this Letter, we present an analytical study of phase structure and critical point of the quantum Euclidean Einstein-Cartan gravity. For the order phase,  calculations and discussions are based on the approaches of strong coupling expansions, the mean-field approximation, and the dynamical equations for holonomy fields. For the disorder phase, we have not been able so far to do analytical calculations in weak gauge couplings region, the discussions on this phase are based on the limit case of gauge coupling $g\rightarrow 0$. The possible ultra-violet critical point and correlation length for the second-order phase transition are estimated in a simple model, according to the competition of the activation energy of holonomy fields and their entropy. Therefore, these results and discussions on the order- and disorder-phase structure,  the ultra-violet critical point and correlation length for the second-order phase transition are preliminary. Numerical simulations are essentially required to check these preliminary results on the phase structure, ultra-violet critical point and correlation length for the second-order phase transition before one can conclude that a sensible continuum field theory of the quantum Euclidean Einstein-Cartan gravity can be defined. 
The coherent correlation length $\xi$ (\ref{loop-par3}) or (\ref{xivary}) is an intrinsic scale of the quantum gravity, analogously to the intrinsic scale $\Lambda_{\rm QCD}$ of the quantum chromodynamics $SU(3)$ theory (QCD) for the strong interaction.  We speculate that the scale $1/\xi^2$ might have some relation to the cosmological constant $\Lambda_{\rm COS}$, and we leave this topic to a further work.  


\noindent
{\bf Acknowledgment.}
\hskip0.1cm
The author thanks H.~W.~Hamber 
for discussions on the Regge calculus and the renormalization group invariance. The author thanks
H.~Kleinert for the discussions on the second-order phase transition. The author also thanks G.~Preparata and R.~Ruffini for discussions on general relativity and Wheeler's foam \cite{wheeler1964}, which brought the author's attention to this topic.

\end{document}